# High-temperature superconductivity on the verge of a structural instability in lanthanum superhydride

Dan Sun[1*], Vasily S. Minkov[2*], Shirin Mozaffari[3], Stella Chariton[4], Vitali B. Prakapenka[4], Mikhail I. Eremets[2], Luis Balicas[3], & Fedor F. Balakirev[1]

1 National High Magnetic Field Laboratory, Los Alamos National Laboratory, NM, US
2 Max-Planck Institut für Chemie, Mainz, Germany
3 National High Magnetic Field Laboratory, Florida State University, Tallahassee, FL, US
4 Center for Advanced Radiation Sources, University of Chicago, Chicago, IL, USA

**A possibility of high, room-temperature superconductivity was predicted for metallic hydrogen in the 1960s. However, metallization and superconductivity of hydrogen are yet to be unambiguously demonstrated in the laboratory and may require pressures as high as 5 million atmospheres. Rare earth based "superhydrides" such as $LaH_{10}$ can be considered a close approximation of metallic hydrogen even though they form at moderately lower pressures. In superhydrides the predominance of H-H metallic bonds and high superconducting transition temperatures bear the hallmarks of metallic hydrogen. Still, experimental studies revealing the key factors controlling their superconductivity are scarce. Here, we report on the pressure and magnetic field response of the superconducting order observed in $LaH_{10}$. For $LaH_{10}$ we find a correlation between superconductivity and a structural instability, strongly affecting the lattice vibrations responsible for the superconductivity.**

For phonon-mediated superconductors, a high transition temperature necessitates light atomic masses. The lightest atom available to compose a lattice is hydrogen, which forms covalently bonded molecular dimmers in ambient conditions. Transforming pure molecular hydrogen with the aid of pressure into a metal with an atomic lattice and into a superconductor has been a long-standing challenge and the subject of contention for the high-pressure community. Yet, chemical pre-compression with certain elements reduces the pressure required for metallization; thus, stable hydrogen-rich phases can be synthesized by the current high-pressure technology. With the discovery of a superconducting transition at the critical temperature $T_c$ = 203 K in $H_3S$[1], the search for hydrogen-rich high-temperature superconductors (HTS) has intensified. A new family of rare-earth hydrides, such as $LaH_{10}$[2, 3] and $YH_9$[4], opened a path for a significant increase in $T_c$ that now approaches room temperature.

While in $H_3S$ the crystal lattice is formed by H-S covalent bonds, $LaH_{10}$ forms a clathrate-like structure, where each La atom is locked in the center of $H_{32}$ hydrogen cage. The interatomic distance between hydrogen atoms in $LaH_{10}$ is close to the H–H distance predicted for atomic metallic hydrogen near $p$ = 500 GPa[5]. Due to the short H–H distance and the high hydrogen content, $LaH_{10}$ can be viewed as 'doped' metallic hydrogen. A pronounced isotope effect on $T_c$ when hydrogen is substituted by its heavier isotope deuterium confirmed that the superconductivity in HTS hydrides is induced by electron-phonon interactions[6]. However, there is a dearth of the experimental studies of HTS hydrides due to a very limited number of measurement techniques available at such extreme pressures. Here we explore the features of the lanthanum hydride family over a wide-range of pressures, temperatures, and magnetic fields to better understand the interplay between its structural and superconducting properties. We find that superconductivity in $LaH_{10}$ is strongly affected by a crystal lattice instability towards symmetry-lowering distortions, a behavior reminiscent of another HTS hydride, namely $H_3S$, where a dramatic change in the $T_c(p)$ dependence is observed at the structural phase transition[7, 8, 9, 10].

Metallic lanthanum reacts readily with excess hydrogen at high pressures and high temperatures yielding clathrate-like superhydrides. We found that the superconducting phase of $LaH_{10}$ can be prepared from the laser-heated mixture of La and $H_2$ at a pressure much lower than 150 GPa[2] or 170 GPa[3, 11] as reported earlier. Specifically, the X-ray powder diffraction data show that the sample synthesized at 138 GPa for this study is well crystallized and comprised of the dominant *Fm-3m* $LaH_{10}$ phase with minor impurities. The impurity phases are attributed to two hexagonal close-packed (*hcp*) phases with *P6$_3$/mmc* crystal lattice symmetry and stoichiometry close to $LaH_{10}$, but with a different *c/a* ratio: ~1.63 for *hcp-I* and ~1.48 for *hcp-II* (Fig. 1). Both impurity phases were found previously in various samples prepared via the direct chemical reaction between hydrogen and lanthanum or lanthanum trihydride[2]. The presence of different amounts of *hcp-I* and *hcp-II* phases in samples at ~140-165 GPa does not distinctly affect the $T_c$ of the superconducting *Fm-3m* phase which has the highest $T_c$ in the lanthanum-hydrogen system[2].

Our sample exhibits a narrow superconducting transition towards zero resistance with a high $T_c$ of 243 K at 138 GPa, slightly lower than the maximum $T_c$ of ~252 K reported for $LaH_{10}$ at ~165 GPa[2,3], in accordance with a "dome-shape" pressure dependence of $T_c$ in the *Fm-3m* phase of $LaH_{10}$[2]. The resistivity $\rho$ of $LaH_{10}$ is estimated to be (0.3 ±0.1) mΩ·cm at $T$ = 300 K and is higher than the value reported for $H_3S$[12]. The large error is mainly due to the uncertainty of the thickness of the sample.

After a decompression from 138 to 120 GPa some reflections from the ancestral cubic phase became split (Fig. 1) and the $T_c$ dropped to 191 K (Fig. 2). The X-ray diffraction powder patterns of the new distorted phase can be reasonably indexed in the *C2/m* space group (Fig. 1b, Supplementary Fig. S1). The refined cell parameters and coordinates of the heavier La atoms are in a good agreement with the theoretical prediction for a monoclinic distortion in $LaH_{10}$ at lower pressures[13, 14]. Recent calculations of the lowest-enthalpy structures for $LaH_{10}$ as a function of pressure found the *R-3m* - to *C2* - to *P-1* sequence as a scenario for lattice distortions[15]. The structural model of the *C2* symmetry assumes a subtle shift of the La atoms out from the mirror plane within the $C2/m$ model, which is about five times smaller than the uncertainty of the refined coordinates for La atom and does not manifest in the experimental X-ray powder patterns.

These structural distortions are reversible, and the $Fm\text{-}3m$ phase of $LaH_{10}$ can be restored if the pressure is increased again. The $T_c$ increases gradually with increasing pressure and reaches 241 K at 136 GPa (Figure 2a). We note that the broadening of the superconducting phase transition in Fig. 2a is likely caused by the deterioration of the phase crystallinity during variations of the pressure. However, the stoichiometry is preserved during the phase transition because the volume expansion is in good agreement with the Equation of State (EoS) for $LaH_{10}$[2, 11]. This observation is also in accordance with the calculations, which indicate that the stoichiometry should not change during any crystal distortion in $LaH_{10}$[13, 14, 15].

The present structural data disagree with the rhombohedral structural distortion scenario towards the $R\text{-}3m$ phase of $LaH_{10}$ at 152 GPa as proposed earlier[11]. Geballe *et al.* studied the structural stability of the $Fm\text{-}3m$ phase of $LaH_{10}$ upon decompression from 169 to 27 GPa and found a phase transition that breaks the lattice symmetry into the $R\text{-}3m$ phase observed between 152 and 121 GPa, followed by decomposition into a compound with a stoichiometry close to $LaH_7$ observed from 109 to 92 GPa. With neither Rietveld nor Le Bail refinements of X-ray diffraction patterns provided, the accuracy of the proposed $R\text{-}3m$ structural model is questionable. The indexing of the powder diffraction patterns was also aggravated by the poor crystallinity of the sample and the high anisotropic stress induced by the lack of a hydrostatic medium inside the diamond anvil cell (DAC) after heating and subsequent absorption of hydrogen by the tungsten gasket. The uncertainty in pressure estimated using different pressure-markers was as high as 30 GPa, indicating large pressure gradients in samples. Thus, the pressure values were systematically overestimated by 15-20 GPa in the previous study[11]. Indeed, the lattice volume per La atom in $Fm\text{-}3m$ $LaH_{10}$ at 138 GPa (34.3 Å$^3$) and $C2/m$ $LaH_{10}$ at 120 GPa (35.4 Å$^3$) refined in the present study corresponds to the same volume values in Ref. 11 at 154 GPa and 139 GPa, respectively. Since the fine crystalline sample synthesized here is surrounded by excess hydrogen, which provides a quasi-hydrostatic medium, and the reflections in the X-ray powder pattern are spotty and narrow, the pressure estimate should be more accurate in the present study. Thus, the pressure value of 152 GPa claimed for the beginning of structural distortions in the $Fm\text{-}3m$ phase of $LaH_{10}$ in Ref. 11 should be reduced to 135 GPa.

The key parameters of the superconducting phase, including the upper critical field, $H_{c2}$, and the superconducting coherence length, $\xi$, for $Fm\text{-}3m$ and $C2/m$ $LaH_{10}$ were determined through magnetotransport measurements at the National High Magnetic Field Laboratory. The samples were electrically connected in a van der Pauw configuration (Fig. 2c, inset), making the measurements of both resistivity and Hall effect possible. The $LaH_{10}$ sample under 120 GPa was measured up to 45 T in DC magnetic fields, and the $LaH_{10}$ sample under 136 GPa was measured in a 65 T pulsed magnet.

The magnetoresistance (MR) of $LaH_{10}$ collected at fixed temperatures is shown in Fig. 3. Under a magnetic field the superconducting transitions span over tens of teslas, which correlates with the broadening of the superconducting transition at zero field (Fig. 2a). The normal state MR above $H_{c2}$ is nearly field- and temperature-independent, with a clear kink at the onset of superconductivity at $H_{c2}$. For consistency with the prior studies, the $H_{c2}$'s are taken as the intersection between the straight-line extrapolations of the normal state magnetoresistance and the slope of the superconducting transition in a method similar to the one followed in Ref. 12. The irreversibility field of the high-temperature superconducting phase ($H^*$) is taken by extrapolating the leading edge of the transition to the horizontal axis (Supplementary Fig. 2). The Hall resistance signal measured above $T_c$ is consistent with the electron-like Fermi surface (Supplementary Fig. 3).

We find that the pressure dependence of $T_c$ in Fig. 2 displays two distinct regions – a low-pressure region characterized by a sharp rise in $T_c$, and a high-pressure region with a much more moderate dome-like $T_c(p)$ dependence, with a clear boundary between the two regions at 135 GPa. This distinct shape in $T_c(p)$ in LaH$_{10}$ closely resembles the $T_c$ variation first discovered in the hydride H$_3$S, and it is explained by the change of the crystalline structure[7, 9, 10]. At high pressures, the H atoms in H$_3$S occupy symmetric positions between S atoms (*Im-3m* symmetry group), but the H sub-lattice distorts slightly in the *R3m* space group as the pressure is decreased. A sharp, but continuous drop in $T_c$ is observed as H$_3$S undergoes a structural transition. Multiple distorted hydrogen arrangements from a high-symmetry *Fm-3m* phase are predicted for LaH$_{10}$ as well[15]. One of the predictions reports a stable LaH$_{10}$ *Fm-3m* phase at high pressures, with symmetric H positions and a $T_c$ of 259 K at 170 GPa. The drop in pressure is predicted to stabilize a distorted *R-3m* phase of LaH$_{10}$, with $T_c$ = 203 K at 150 GPa[14]. A $T_c$ ~ 229-245 K was calculated for the *C2/m* phase, although the calculations were performed for $p$ = 200 GPa, which is substantially higher than the values presented here[13].

A likely explanation for the sharp change in the dependence of $T_c$ with pressure below 135 GPa, is a structural phase transition in LaH$_{10}$. The lack of a discontinuous jump in $T_c$ in LaH$_{10}$ and in H$_3$S[7, 9] points to a continuous symmetry-lowering lattice distortion or a phase transition of the second-order. Although a possible first-order transition was also proposed for H$_3$S[10], it would result in a discontinuous jump in $T_c$, which we do not observe.

A higher-symmetry crystal structure transformation into a lower-symmetry phase is governed by phonon softening when the frequency of the collective atomic movement approaches zero. Such a drastic change in phonon modes often has a profound effect on the phonon-mediated superconducting order. A boost in $T_c$ due to phonon softening in the vicinity of a structural transition has been reported in a number of superconducting families, ranging from Sn nanostructures[16], A15 compounds[17], intercalated graphite[18], ternary silicides[19], and even some elements under pressure[20, 21]. A symmetry-lowering distortion in the H sub-lattice in LaH$_{10}$ is driven by softening of some of the H-H phonon modes, leading to stronger electron-phonon interaction in the *Fm-3m* phase, which is characterized by a coupling constant $\lambda = 2\int_0^\infty \alpha^2 F(\omega)\omega^{-1}d\omega$, where $\omega$ is phonon frequency, $F(\omega)$ is the phonon density of states, and $\alpha^2$ is an average square electron-phonon matrix element. While the light atomic mass of hydrogen is a necessary requirement for phonon-coupled HTS, the $T_c$ is also strongly affected by $\lambda$ [22, 23], with a peak in $T_c$ predicted for large $\lambda$ ~ 2-2.5, which should occur in the close proximity to lattice instability in HTS hydrides[24].

Upper critical field measurements in $H_3S$ HTS hydride have independently verified a large $\lambda \sim 2$[12]. We find a substantially larger $H_{c2}$ for $LaH_{10}$ and determined that magnetic fields of the order of 100 T will be required to distinguish between a strongly-coupled scenario with a large $\lambda$ and more commonly employed Werthamer-Helfand-Hohenberg (WHH) model derived in a weakly-coupled limit, $\lambda \ll 1$[25]. To extract the key superconducting properties of $LaH_{10}$ and explore the effects of the structural transition on superconductivity, we fit the temperature dependence of $H_{c2}$ to WHH (Figure 4 and Table 1). WHH model fits well with our data up to 60 T, our upper measurement limit. The WHH model considers the combined effects of magnetic field on the orbital motion and on the spin of electrons: $H_{c2}^{-2} = H_{c\,orb}^{-2} + H_{c\,p}^{-2}$, where $H_{c\,orb}$ and $H_{c\,p}$ are the orbitally-limited and spin-limited (Pauli) critical fields, respectively. We obtain $H_{c\,p}(0)$ values of 352 T at 120 GPa and 457 T at 136 GPa. $H_{c\,p}(0)$ values are by a factor of ~3 larger than $H_{c2}(0)$ values listed in Table 1, indicating predominantly orbital-limited upper critical field in HTS $LaH_{10}$, analogous to $H_3S$[12].

The WHH fit provides a reasonable estimate of the superconducting coherence length $\xi = \sqrt{\phi_0/2\pi H_{c2}}$, where $\phi_0$ is the magnetic flux quantum. There is a significant drop in $T_c$ in the distorted phase of $LaH_{10}$ at 120 GPa when compared to the $LaH_{10}$ sample at 136 GPa. Surprisingly, $H_{c2}(0)$ only drops by a small amount and, thus, $\xi(0)$ remains nearly unchanged. $\xi$ is linked to both $T_c$ and the Femi velocity $v_F$: $\xi = 0.18\,\hbar v_F / k_B T_c$ within BCS theory[6], but the $\xi \sim v_F/T_c$ rule should remain valid for other models, thus signaling a decreased $v_F$ in the *C2/m* phase at 120 GPa when compared to that *Fm-3m* phase at 136 GPa. The onset of the lattice distortion is expected to strongly affect the electron dispersion, e.g. via the flattening of the bands at the new Brillouin zone boundaries, which may lead to a drop in $v_F$ so that $\xi$ and $H_{c2}$ remain high despite the drop in $T_c$ in *C2/m* phase.

In conclusion, we have measured the properties of $LaH_{10}$ HTS superhydride as a function of pressure, temperature, and high magnetic fields. We find evidence for a pressure-induced structural transition in $LaH_{10}$ at $p_c = 135$ GPa, resulting in a steep, but continuous decrease in $T_c(p)$ below $p_c$. A likely mechanism for the structural instability is phonon softening and a gradual distortion of the lattice, in common with another HTS hydride $H_3S$. We established key superconducting quantities of superhydrides under high magnetic fields, including upper critical fields and coherence lengths. We find that the drop in the Femi velocity in $LaH_{10}$ is consistent with the Brillouin zone changes induced by the distortions. The proximity of peak $T_c$ and symmetry-lowering structural transition, which is now experimentally established for at least two HTS hydride families, indicates that tuning of the soft phonon modes should be viewed as one of the main pathways for maximizing $T_c$ in hydride superconductors.

**ACKNOWLEDGMENTS**

The work performed at the National High Magnetic Field Laboratory is supported by the National Science Foundation Cooperative Agreement No. DMR-1644779, and the State of Florida. L. B. is supported by the Department of Energy, Basic Energy Sciences through award DE-SC0002613. The synchrotron X-Ray diffraction data were collected at GeoSoilEnviro CARS (The University of Chicago, Sector 13), Advanced Photon Source (APS), Argonne National Laboratory (USA).

GeoSoilEnviro CARS is supported by the National Science Foundation-Earth Sciences (EAR-1634415) and Department of Energy-GeoSciences (DE-FG02-94ER14466). This research used resources of the Advanced Photon Source, a U.S. Department of Energy (DOE) Office of Science User Facility operated for the DOE Office of Science by Argonne National Laboratory under Contract No. DE-AC02-06CH11357. M.I.E acknowledge great support from Max Planck Society.

**AUTHOR CONTRIBUTION**

D.S., V.S.M. and F.F.B. designed the research and wrote the paper; V.S.M. prepared the samples, collected synchrotron X-Ray diffraction data, performed electrical transport measurements without external magnetic field and processed the structural data; D.S., F.F.B., S.M., L.B. and M.I.E. performed electrical transport measurements under external magnetic fields and processed the data; D.S., F.F.B., S.C. and V.B.P. assisted with the synchrotron X-ray diffraction experiments; M.I.E. designed the diamond anvil cell. All authors contributed to writing the paper. D.S. and V.S.M. contributed equally to this work.

**DATA AVAILABILITY STATEMENT**

The data that support the findings of this study are available from the corresponding author upon reasonable request.

**ADDITIONAL INFORMATION**

Supplementary Information is available for this paper. Correspondence and requests for materials should be addressed to D.S. and F.F.B..

Table 1. Summary of samples properties and the associated WHH fit parameters: the critical temperature, the upper critical field at $T=0$, coherence length at $T=0$, BCS Fermi velocity, and the slope of $H_{c2}$ at the critical temperature.

| Sample structure | $p$ (GPa) | $T_c$ (K) | $H_{c2}(T=0)$ (T) | $\xi(T=0)$ (nm) | $v_F$ ($\times 10^5$ m/s) | $dH_{c2}/dT|_{Tc}$ (T/K) |
|---|---|---|---|---|---|---|
| $C2/m$ LaH$_{10}$ | 120 | 189 | 133.5 | 1.57 | 2.17 | -1.12 |
| $Fm\text{-}3m$ LaH$_{10}$ | 136 | 246 | 143.5 | 1.514 | 2.77 | -0.83 |

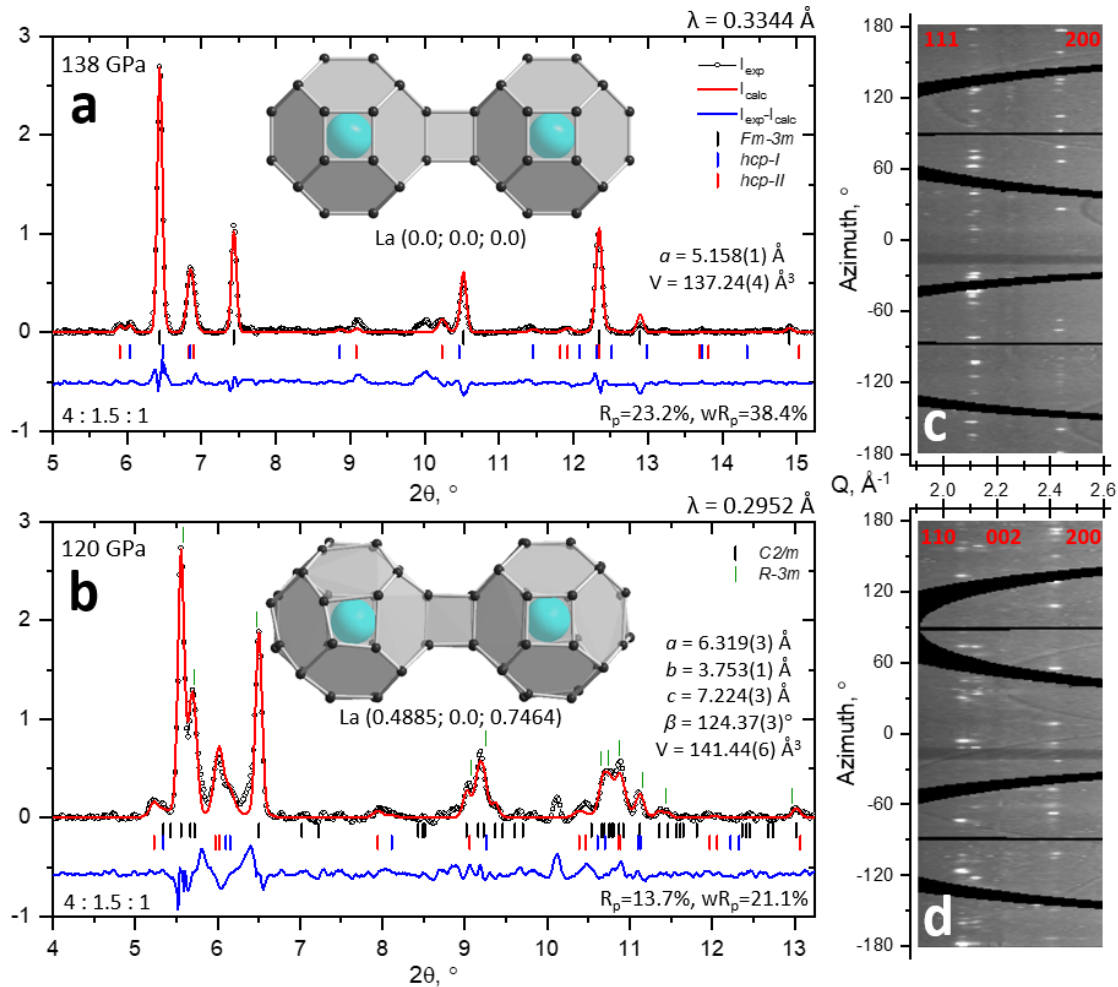

Figure 1. Structural data for LaH$_{10}$ synthesized from La and excess H$_2$. **a**, **b**, Rietveld refinement for *Fm-3m* phase of LaH$_{10}$ at 138 GPa and *C2/m* phase of LaH$_{10}$ at 120 GPa, respectively. The peaks originating from the *hcp-I* ($a$=3.668(4) Å; $c$=5.914(11) Å; $V$=68.9(1) Å$^3$ at 138 GPa) and *hcp-II* ($a$=3.750(3) Å; $c$=5.561(7) Å; $V$=67.7(1) Å$^3$ at 138 GPa) impurity phases having *P6$_3$/mmc* symmetry and LaH$_{10}$ stoichiometry are indicated through blue and red dashes, respectively. Green dashes corresponding to reflection positions in the *R-3m* structural model ($a$=3.73(1) Å; $c$=8.89(1) Å; $V$=107.2(2) Å$^3$), hardly fit the experimental powder pattern and are significantly shifted from the observed maxima (Supplementary Fig. S1). The refined ratio between the main and the impurity phases is provided in the left bottom corner of each figure. The main structural building block, two connected LaH$_{32}$ polyhedra, for each phase are shown in the middle inserts. Large blue and small black spheres correspond to La and H atoms, respectively. H atoms were placed in the calculated positions using predicted structural models[13]. **c**, **d**, The original X-ray diffraction powder patterns at 138 GPa and 120 GPa, respectively. New reflections appear at 120 GPa due to the monoclinic distortions. These reflections are absent in *Fm-3m* lattice at 138 GPa.

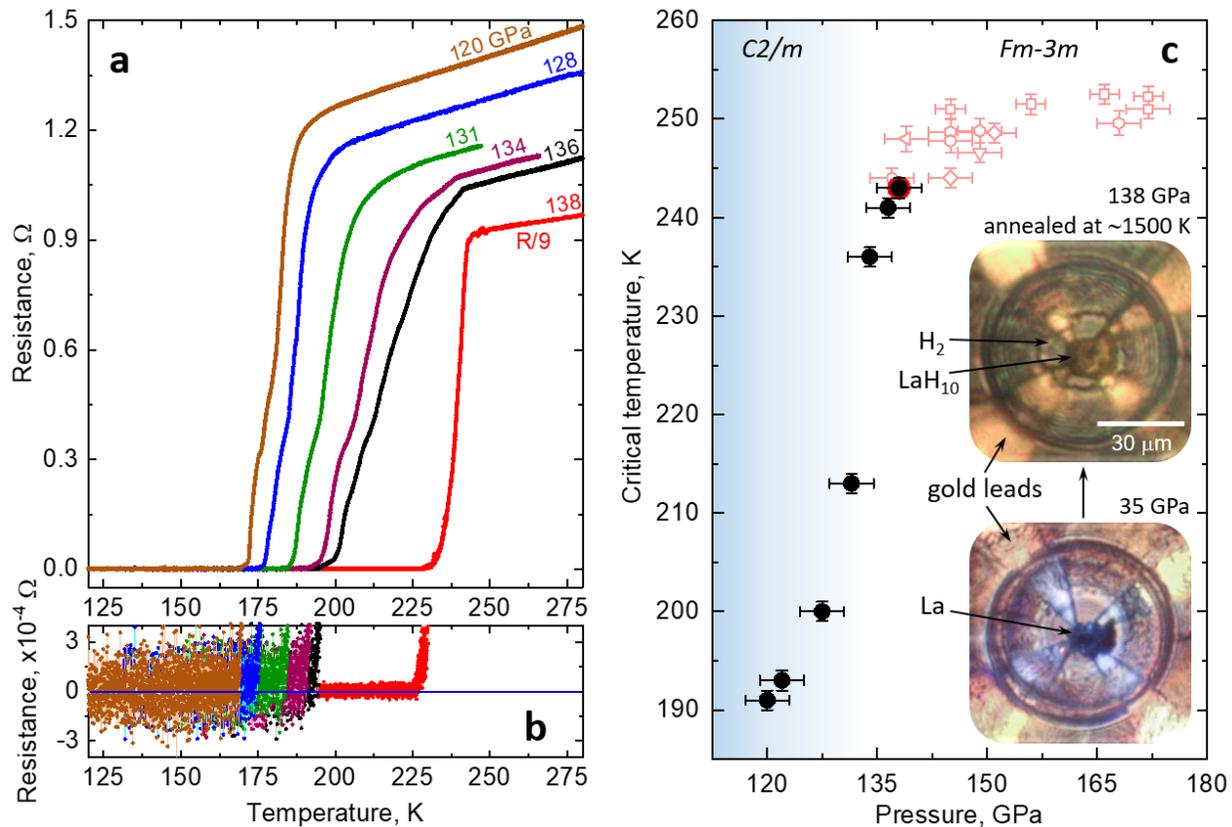

Figure 2. The superconducting transitions in $LaH_{10}$. **a**, **b**, The electrical resistance in $LaH_{10}$ after the synthesis at 138 GPa (red curve), after decompression down to 120 GPa (brown curve), and upon a gradual increase in pressure from 120 to 136 GPa (blue, green, purple, and black curves). The data measured at 138 GPa are divided by 9 for better presentation. **c**, Pressure dependence of $T_c$ in $LaH_{10}$ measured in the present study (black symbols), and from a prior study[2] (open red symbols). Inserts: photos of the DAC loaded with a La flake and after the synthesis of $LaH_{10}$ through laser-assisted heating.

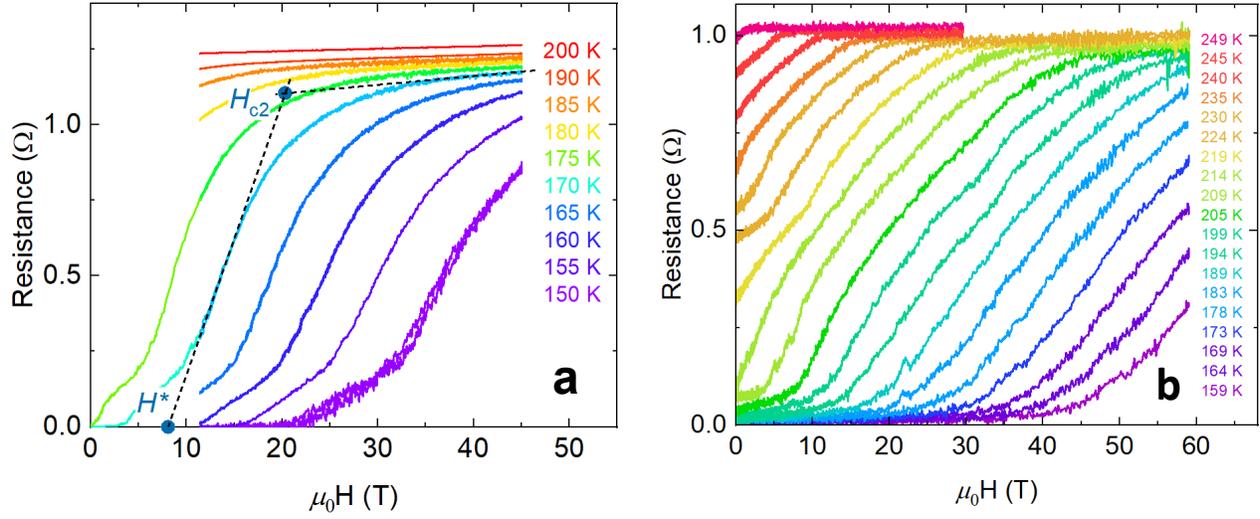

Figure 3. Field dependence of the resistance of LaH$_{10}$ at different temperatures. **a**, DC field measurements for the *C2/m* phase of LaH$_{10}$ at 120 GPa. Two dashed lines extrapolate the slope of the high-temperature superconducting transition (left line) towards the asymptotic trace representing the high field normal state magnetoresistance (right line) at 170 K, respectively. The intersection between two lines provides an estimation of the upper critical field ($H_{c2}$). The intersection of the first line with horizontal axis indicates the irreversibility field (H*) for the high temperature superconducting phase. **b**, Pulsed field measurements for the *Fm-3m* phase of LaH$_{10}$ at 136 GPa. Both DC and pulsed field traces were recorded under isothermal conditions and no eddy-current generated Joule heating due to the sweeping of the field was detected.

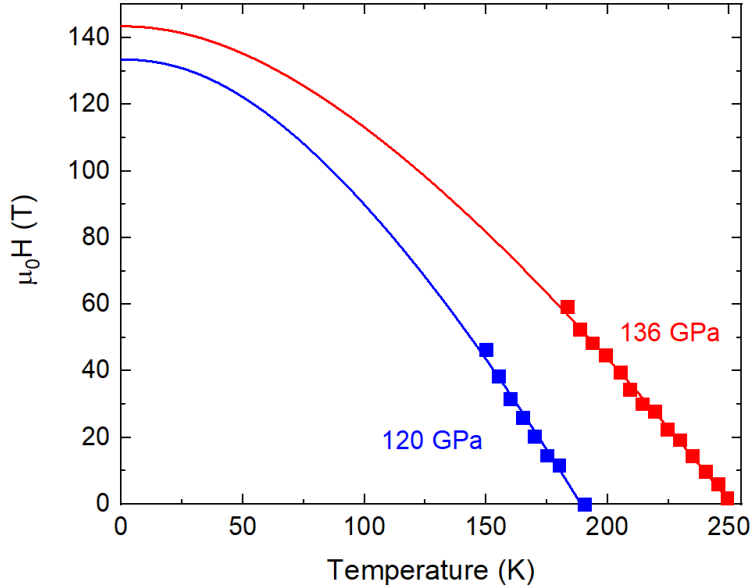

Figure 4. Fits of the superconducting upper critical $H_{c2}$ to the Werthamer-Helfand-Hohenberg (WHH) formalism. Red and blue squares denote the loci of $H_{c2}$ of $LaH_{10}$ at 136 GPa and 120 GPa, respectively. Lines of the same color are the WHH fits to the experimental data.

## Methods

**Diamond anvil cell**

The samples of $LaH_{10}$ were synthesized *in situ* in specially designed miniature diamond anvil cells with a maximum diameter of 8.8 mm and a body length of ~30 mm[1]. The DAC is small enough to fit in the narrow bore of high field DC and pulsed magnets, and is still able to provide a wide-angle optical opening for laser-assisted synthesis and X-ray measurements, and to reach ultra-high pressures up to 200 GPa.

**Sample preparation**

For the sample synthesis, a small piece of metallic lanthanum (Alfa Aesar, 99.9%) with a lateral dimension of about 10 μm and a thickness of ~1-2 μm was placed in the center of the beveled diamond anvil with a culet size of 35 μm onto the tips of four sputtered leads. Sputtered gold electrodes were thoroughly isolated from the metal rhenium gasket by a protecting layer made from magnesium oxide, calcium fluoride, and epoxy glue mixture. Excess hydrogen ($H_2$, 99.999%; $D_2$, 99.75%) was introduced in the DAC at a gas pressure of about 150 MPa. After the cell was thoroughly clamped, the sample was pressurized to the desired pressures (138 GPa for La-H) and then heated up to ~1500-2000 K by a microsecond pulse YAG laser to initiate the chemical reaction between reactants. The pressure was estimated from the Raman shift of the stressed diamond edge[26] and the vibron of $H_2$[27]. Both scales indicated the same pressure within an error of ±5 GPa.

**Structure characterization**

X-ray diffraction data were collected at the beamline 13-IDD at GSECARS, Advanced Photon Source using $\lambda_1$=0.2952 Å and $\lambda_2$=0.3344 Å, beam spot size of ~3 x 3 μm, and Pilatus 1M CdTe detector. Typical exposure time varied between 10-300 s. Processing and integration of the X-ray diffraction powder patterns were carried out using the Dioptas software[28]. Indexing and Rietveld refinement were performed in GSAS and EXPGUI packages[29, 30]. The coordinates of the heavier lanthanum atoms were refined, whereas H atoms were placed in the calculated positions derived from the theoretically proposed models[13, 15, 31]. Structural data for the refined phases of $LaH_{10}$ can be obtained as Crystallographic Information Files from the Cambridge Crystallographic Data Centre via www.ccdc.cam.ac.uk/data_request/cif, on quoting the Deposition Number: 2033292-2033293

**Magnetotransport measurements**

Zero field electrical resistance was measured through a four-probe technique in van der Pauw geometry with currents ranging from $10^{-4}$ A at $p$ = 138 GPa to $10^{-3}$ A at $p$ = 120 - 136 GPa samples. No apparent effect of the current value on the measured $T_c$ was observed. The electrical measurements were obtained in a warming part of a thermal cycle as it yields a more accurate temperature reading: a sample is warmed up slowly (0.2 K min$^{-1}$) under nearly isothermal environmental conditions (no coolant flow). The temperature was measured by a Si diode thermometer attached to the DAC with an accuracy of ~0.1 K. $T_c$ was determined at the offset of superconductivity – at the point of apparent deviation in the temperature dependence of the resistance from the normal metallic behavior. The magnetotransport under high magnetic fields was measured in 45T hybrid magnet and in 65 T pulsed magnet at the National High Magnetic Field Laboratory. A copper thermal shield was placed around the DAC during DC field measurements. The thermal shield was heated uniformly to reduce the thermal gradients, and a secondary Cernox thermometer was attached to the DAC gasket for accurate measurements of the sample temperature. There is no observable heating from the ramping of the magnetic field at rates up to 3 T/min. The Hall effect was measured for the sample at 120 GPa above $T_c$ in hybrid DC magnet from 11.5 T to 45 T. Reverse-field reciprocity method was employed to determine Hall resistance $R_{xy}$[32] because the field direction of the hybrid magnet cannot be reversed during the day shift. A high-frequency (290 kHz) lock-in amplifier technique was employed to measure sample magnetoresistance in 65 T pulsed magnet. 500 μA AC current was applied to the sample, the voltage drop across the sample was amplified by an instrumentation amplifier and detected by a lock-in. No sample heating was observed during ~50 ms long magnet pulse based on comparisons of up sweep and down sweep resistance traces at different field sweep rates.

**Werthamer-Helfand-Hohenberg model**

Numerical fit to the Werthamer-Helfand-Hohenberg model for the temperature dependence of $H_{c2}(T)$ defined by orbital and spin-paramagnetic effects in the dirty limit is given by WHH[25]:

$$\ln\left(\frac{1}{t}\right) = \sum_{\nu=-\infty}^{\infty}\left\{\frac{1}{|2\nu+1|} - \left[|2\nu+1| + \frac{\bar{h}}{t} + \frac{(\alpha\bar{h}/t)^2}{|2\nu+1| + (\bar{h}+\lambda_{so})/t}\right]^{-1}\right\}$$

where $h = (4/\pi^2)[H_{c2}(T)/T_c(-dH_{c2}/dT)_{Tc}]$, $\alpha$ is the Maki parameter, and $\lambda_{SO}$ is the spin-orbit constant. The Maki parameter for each sample is estimated from the slope of $H_{c2}(T)$ at $T=T_c$: $\alpha = \sqrt{2}\, H_{c\,orb}/H_{c\,p} \sim -0.52758\, dH_{c2}/dT|_{Tc}$ [25].

# Supplementary Information for

# "High-temperature superconductivity on the verge of structural instability in lanthanum superhydride"

Dan Sun, Vasily S. Minkov, Shirin Mozaffari, Stella Chariton, Vitali B. Prakapenka, Mikhail I. Eremets, Luis Balicas & Fedor F. Balakirev

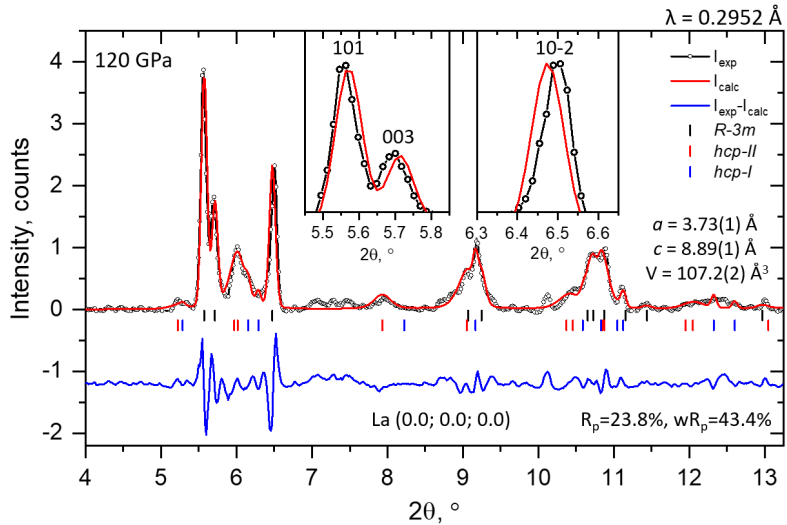

Supplementary Figure 1. The alternative Rietveld refinement for *R-3m* phase of LaH$_{10}$ at 120 GPa, which demonstrates the worse fitting for the experimental X-ray diffraction powder pattern in comparison with the *C2/m* structural model. The enlarged inserts show the significant shifts between the calculated positions for the first three 101, 003, and 10-2 reflections in the *R-3m* structural model from the observed maxima. The refined fitting factors are $R_p$=23.8% and $wR_p$=43.4% and considerably higher than for *C2/m* model ($R_p$=13.7%, $wR_p$=21.1%).

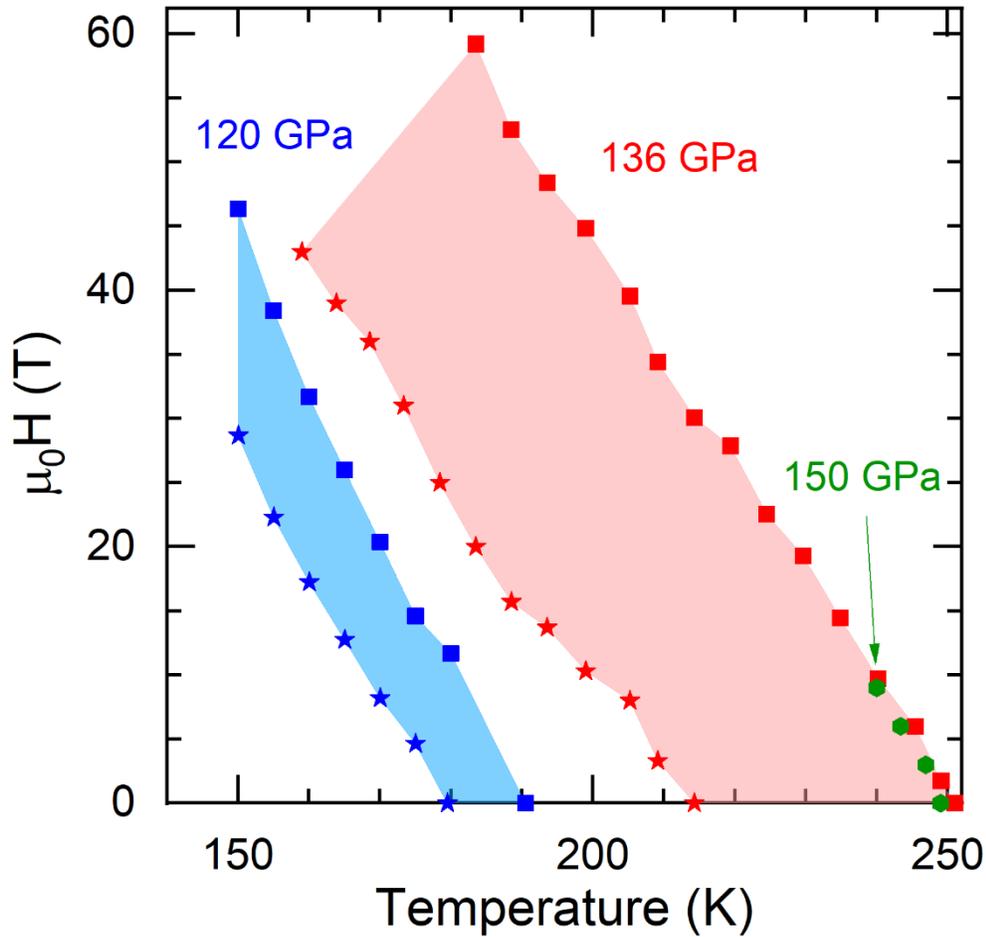

Supplementary Figure 2. The upper critical field $H_{c2}$ and the vortex melting field $H^*$ in LaH$_{10}$. $H^*$ is estimated by extrapolating the leading edge of the transition to horizontal axis and $H_{c2}$ at the offset from the normal state magnetoresistance in $R(H)$ traces form Fig. 3. Stars denote the loci of $H^*$, and squares denote that of $H_{c2}$. The blue and red symbols are for LaH$_{10}$ at 120 GPa and 136 GPa, respectively. The green hexagons are the $H_{c2}$ from reference[1]. The region of resistive dissipation between $H^*$ and $H_{c2}$ is the so-called vortex liquid state, denoted with red and blue shades. The measurement at 120 GPa has a narrower vortex liquid region while this region becomes wide after pressure is increased to 136 GPa. The upper critical field $H_{c2}$ of the 136 GPa measurement agrees well with the previous measurement up to 9 T in a sample with similar $T_c$[1].

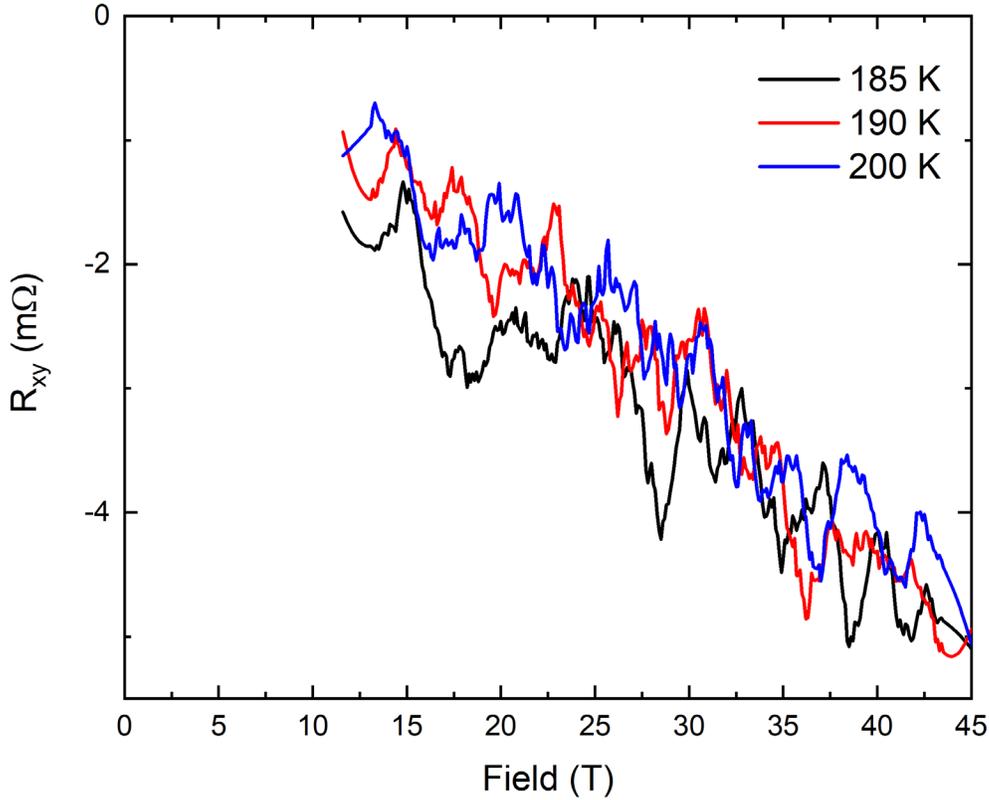

Supplementary Figure 3. The Hall resistance in the LaH$_{10}$ sample at 120 GPa. The Hall effect is measured above $T_c$ in a DC magnet by sweeping the resistive insert field between 11.5 T and 45 T. $R_{xy}$ is linear at all three temperatures, which indicates that the magnetotransport in LaH$_{10}$ above $T_c$ is dominated by a single electron band. The Hall coefficient $R_H$ value is obtained by taking the slope of $R_{xy}$ and multiplying it by the thickness of the sample. The three measurements yield values ~ $2.3 \times 10^{-10} m^3/C$. Compared with $0.7 \times 10^{-10} m^3/C$ for H$_3$S$^2$, this $R_H$ value for LaH$_{10}$ is 3 times larger. From a simple single band model, the density of electrons is $n = -1/R_H e = 7.2 \times 10^{21} cm^{-3}$. This value would indicate that LaH$_{10}$ has lower carrier density than H$_3$S, if one disregards some uncertainty about the actual conductive path within the sample. The temperature variance at this temperature range is small, although the variance may come from the noise in the measurement. We find that the Hall signal is lower than the noise in data collected under pulsed-fields at a pressure of 136 GPa, and thus it is not presented here.